\documentclass[12pt]{article}
\usepackage{amssymb,amsmath}
\usepackage{epsfig}

\renewcommand{\theequation}{\arabic{section}.\arabic{equation}}

\def\N{{\mathcal N}}
\def\L{{\mathcal L}}

\def\L{{\mathcal L}}

\def\ds{\displaystyle}

\def\a{\alpha}
\def\b{\beta}

\def\g{\gamma}

\def\s{\sigma}
\def\t{\tilde}

\def\m{\mu}

\def\l{\lambda}

\def\rb{\right}
\def\lb{\left}
\def\axs{AdS_5\times S^5}
\newcommand{\eq}[1]{\begin{equation} #1 \end{equation}}
\newcommand{\al}[1]{\begin{align} #1 \end{align}}
\newcommand{\ml}[1]{\begin{multline} #1 \end{multline}}

\begin{document}

\begin{titlepage}

\begin{center}

\vspace*{3cm}

\centerline{\Large \bf Spiky Strings, Giant Magnons and
$\beta$-deformations}

\vspace*{1cm} N.P. Bobev${}^{\star}$\footnote{email:
bobev@usc.edu} and R.C. Rashkov${}^{\dagger}$\footnote{e-mail:
rash@hep.itp.tuwien.ac.at; on leave from Dept of Physics, Sofia
University, Bulgaria.

To Prof. K.S.Viswanathan on the occasion of his 70-th birthday.}

\ \\
${}^{\star}$ \textit{Department of Physics and Astronomy ,
University of Southern California,\\ Los Angeles, CA 90089-0484,
USA}

\ \\
${}^{\dagger}$ \textit{Institute for Theoretical Physics, Vienna
University of Technology,\\
Wiedner Hauptstr. 8-10, 1040 Vienna, Austria}

\end{center}

\vspace*{.8cm}

\baselineskip=18pt

\bigskip
\bigskip
\bigskip
\bigskip

\begin{abstract}

We study rigid string solutions rotating on the $S^3$ subspace of
the $\beta$-deformed $AdS_5\times S^5$ background found by Lunin
and Maldacena. For particular values of the parameters of the
solutions we find the known giant magnon and single spike strings.
We present a single spike string solution on the deformed $S^3$
and find how the deformation affects the dispersion relation. The
possible relation of this string solution to spin chains and the
connection of the solutions on the undeformed $S^3$ to the
sine-Gordon model are briefly discussed.
\end{abstract}

\end{titlepage}

\section{Introduction}

    The idea for the correspondence between the large N limit of gauge
theories and string theory was proposed over thirty years ago
\cite{'tHooft:1973jz} and an explicit realization of it was
provided when Maldacena conjectured the AdS/CFT correspondence
\cite{holography}. Since then this became a major research area
and many fascinating discoveries were made in the last decade.

Recent advances on both the string and the gauge theory sides of
the correspondence have indicated that type IIB string theory on
$\axs$ and $\N = 4$ super-Yang-Mills (SYM) theory in four
dimensions may be integrable in the planar limit. The techniques
of integrable systems have therefore become useful in studying the
AdS/CFT correspondence in detail. One of the conjectures of the
correspondence is the duality between the spectrum of anomalous
dimensions of gauge invariant operators in the the gauge theory
and the energy spectrum of the string theory.

Assuming that these theories are integrable, the dynamics should
be encoded in an appropriate scattering matrix $S$. This can be
interpreted from both sides of the correspondence as follows. On
the string side, in the strong-coupling limit the $S$ matrix can
be interpreted as describing the two-body scattering of elementary
excitations on the worldsheet. When their worldsheet momenta
become large, these excitations can be described as special types
of solitonic solutions, or giant magnons, and the interpolating
region is described by the dynamics of the so-called
near-flat-space regime \cite{Hofman:2006xt,Swanson:2006dec}. On
the gauge theory side, the action of the dilatation operator on
single-trace operators is the same as that of a Hamiltonian acting
on the states of a certain spin chain \cite{Minahan:2002ve}. This
turns out to be of great advantage because one can diagonalize the
matrix of anomalous dimensions by using the algebraic Bethe ansatz
technique (see \cite{fadd} for a nice review on the algebraic
Bethe ansatz). In this picture the dynamics involves
diffractionless scattering encoded by an S matrix. Proving that
the gauge and string theories are identical in the planar limit
therefore amounts to showing that the underlying physics of both
theories is governed by the same two-body scattering matrix. On
the other hand, in several papers the relation between strings and
spin chains was established at the level of effective action, see
for instance
\cite{kruczenski},\cite{dim-rash},\cite{lopez},\cite{tseytlin1}
and references therein. These ideas opened the way for a
remarkable interplay between spin chains, gauge theories, string
theory\footnote{For very nice reviews on the subject with a
complete list of references see
\cite{beisphd},\cite{tseytrev},\cite{Plefka:2005bk}} and
integrability (the integrability of classical strings on $\axs$
was proven in \cite{bpr}).

Recently Hofman and Maldacena \cite{Hofman:2006xt} were able to
map spin chain "magnon" excitations to specific rotating
semiclassical string states on $R\times S^2$. These strings move
around the equator of the $S^2$ and have very large energy and
angular momentum. The momentum of the spin chain magnon was
interpreted as a deficit angle of the string configuration. This
result was soon generalized to magnon bound states
\cite{Dorey1},\cite{Dorey2},\cite{AFZ},\cite{MTT}, dual to strings
on $R\times S^3$ with two non-vanishing angular momenta. Moreover
in \cite{MTT} a different giant magnon state with two spins was
found, it is dual to string moving on $AdS_3\times S^1$ i.e. it
has spin in both the AdS and the spherical part of the background
(see also \cite{Ryang:2006yq}). These classical string solutions
were further generalized to include dynamics on the whole $S^5$
\cite{volovich}, \cite{Kruczenski:2006pk}  and in fact a method to
construct classical string solutions describing superposition of
arbitrary scattering and bound states was found
\cite{Kalousios:2006xy}. The semiclassical quantization of the
giant magnon solution was performed in
\cite{Papathanasiou:2007gd}.

The relation between energy and angular momentum for the one spin
giant magnon found in \cite{Hofman:2006xt} is:
\eq{ E-J=\ds\frac{\sqrt{\l}}{\pi}|\sin\ds\frac{p}{2}| }
where $p$ is the magnon momentum which on the string side is
interpreted as a difference in the angle $\phi$ (see
\cite{Hofman:2006xt} for details). In the multispin cases the
$E-J$ relations were studied both on the string
\cite{Dorey2},\cite{AFZ},\cite{MTT} and spin chain sides
\cite{Dorey1}. A natural way to extend this analysis and find
giant magnon solutions in backgrounds which via the AdS/CFT
correspondence are dual to less supersymmetric gauge theories is
to look for similar giant magnon solutions in the Lunin-Maldacena
background \cite{Chu:2006ae},\cite{Bobev:2006fg}.


A class of classical string solutions with spikes maybe of
interest for the AdS/CFT correspondence. These spiky strings were
constructed in the $AdS_5$ subspace of $AdS_5\times S^5$
\cite{Kruczenski:2004wg} and it was argued that they correspond to
single trace operators with a large number of derivatives. The
spiky strings were generalized to include dynamics in the $S^5$
part of $AdS_5\times S^5$ \cite{Ryang:2005yd} and in
\cite{McLoughlin:2006tz} closed strings with "kinks" were
considered\footnote{For other spiky string solutions see
\cite{Mosaffa:2007}}.

Recently a new analysis of the class of spiky string  appeared, in
\cite{Ishizeki:2007we} infinitely wound string solutions with
single spikes on $S^2$ and $S^3$ were found. It can be shown that
these solutions can be found in a certain limit of the parameters
of a general rotating rigid string. Interestingly enough the giant
magnon solution of \cite{Hofman:2006xt} can be found in a very
similar way and thus the single spike solutions fall into the same
class as the giant magnons. While the interpretation of the giant
magnon solutions from the field theory point of view is as of
higher twist operators, the single spike solutions do not seem to
be directly related to some gauge theory operators. However in
\cite{Ishizeki:2007we} an interpretation of this solution as a
spin chain Habbard model, which means antiferromagnetic phase of
the corresponding spin chain, was found, but the relation to field
theory operators is still unclear.

The single spike solutions have interesting properties, for
instance the energy and ``the angle deficit'' $\Delta\phi$ are
infinite, but their difference is finite
\eq{
E-T\Delta\phi=\frac{\sqrt{\l}}{\pi}\left(\ds\frac{\pi}{2}-\theta_1\right)
\notag }
where $T=\frac{\sqrt{\l}}{2\pi}$ is the string tension. The
angular momenta of the solution are finite and in the case of
single spike solution on $S^3$ their relation
\begin{equation}
J_1 = \sqrt{J_2^2 + \ds\frac{\lambda}{\pi^2} \cos^2\theta_1}
\end{equation}
resembles the dispersion relation of giant magnons on $S^3$. The
single spike solutions have relation to the sine-Gordon model and
this could be useful to study their scattering and eventually lead
to a better understanding of their properties. It seems
interesting to study these single spike solutions in greater
detail and generalize them to less symmetric backgrounds. The
purpose of this paper is to present single spike solutions on the
Lunin-Maldacena \cite{Lunin:2005jy} background which will be the
counterparts of the giant magnon solutions on this background
constructed in \cite{Chu:2006ae}, \cite{Bobev:2006fg}. We find
that the close relation between the single spikes and giant
magnons known from the solutions on $AdS_5\times S^5$ persists in
the deformed background. The relation between the energy and the
angular momenta is similar and the effect of the deformation is
similar to the case of the giant magnon.


The paper is organized as follows. First in Section 2 we present a
short review of the Lunin-Maldacena background, then we discuss
the dynamics of strings on a $S^3$ subspace of the deformed $S^5$.
In Section 3 we consider vanishing deformation parameter and find
classical string solutions on $S^2$ and $S^3$ in conformal gauge
and by taking different limits we reproduce the giant magnon and
single spike solutions found in \cite{Hofman:2006xt} and
\cite{Ishizeki:2007we}. In Section 4 we show the relation of the
single spike and giant magnon solutions on $S^3$ to the
sine-Gordon model. Then in Section 5 we generalize the single
spike solution of \cite{Ishizeki:2007we} to a similar solution on
the $\gamma$-deformed $S^3$ and establish it similarities with the
giant magnons on $S^3_{\gamma}$ found in \cite{Bobev:2006fg}. In
the last section we present our conclusions and a few possible
directions for further studies.

\section{Rigid strings on $S^3_{\gamma}$}

\subsection{$\b$-deformed $\axs$}

Here we review the $\b$-deformed $\axs$ background found by Lunin
and Maldacena \cite{Lunin:2005jy}. This background is conjectured
to be dual to the Leigh-Strassler marginal deformations of $\N=4$
SYM \cite{Leigh:1995ep}. We note that this background can be
obtained from pure $\axs$ by a series of STsTS transformations as
described in \cite{Frolov}. The deformation parameter
$\b=\g+i\sigma_d$ is in general a complex number, but in our
analysis we will consider $\s_d=0$, in this case the deformation
is called $\gamma$-deformation. The resulting supergravity
background dual to real $\b$-deformations of $\N=4$ SYM is:
\eq{
ds^2=R^2\left(ds^2_{AdS_5}+\ds\sum_{i=1}^{3}(d\m_i^2+G\m_i^2d\phi_i^2)+
\t{\g}^2G\m_1^2\m_2^2\m_3^2(\ds\sum_{i=1}^3d\phi_i^2)\right)
\label{3.1}}
This background includes also a dilaton field as well as RR and
NS-NS form fields. The relevant form for our classical string
analysis will be the antisymmetric B-field:
\eq{
B=R^2\t{\g}G\left(\m_1^2\m_2^2d\phi_1d\phi_2+\m_2^2\m_3^2d\phi_2d\phi_3+\m_1^2\m_3^2d\phi_1d\phi_3\right)
\label{3.2}}
In the above formulae we have defined
\eq{\begin{array}{l} \t{\g}=R^2\g \qquad\qquad R^2=\sqrt{4\pi
g_sN}=\sqrt{\l}
\\\\
G=\ds\frac{1}{1+\t{\g}^2(\m_1^2\m_2^2+\m_2^2\m_3^2+\m_1^2\m_3^2)}\\\\
\m_1=\sin\theta\cos\psi \qquad \m_2=\cos\theta \qquad
\m_3=\sin\theta\sin\psi
\end{array}\label{3.3}}
Where ($\theta$,$\psi$,$\phi_1$,$\phi_2$,$\phi_3$) are the usual
$S^5$ variables. This is a deformation of the $AdS_5\times S^5$
background governed by a single real deformation parameter
$\tilde{\gamma}$ and thus provides a useful setting for the
extension of the classical strings/spin chain/gauge theory duality
to less supersymmetric cases.

\subsection{Rigid strings on $S^3_{\gamma}$}

Let us consider the motion of a rigid string on $S^3_{\gamma}$.
This space can be thought of as a subspace of the
$\gamma$-defromation of $AdS_5\times S^5$ presented above
\begin{equation}
\m_3=0, \quad \phi_3=0 \quad \text{i.e.} \quad \psi=0, \quad
\phi_3=0. \label{3-sphere}
\end{equation}
The relevant part of the $\gamma$-deformed $AdS_5\times S^5$ is
\begin{equation}
ds^2 = -dt^2 + d\theta^2 + G\sin^2\theta d\phi_1^2 +G\cos^2\theta
d\phi_2^2 \label{deformedmetric}
\end{equation}
where $G = \ds\frac{1}{1+\tilde{\gamma}^2
\sin^2\theta\cos^2\theta}$ and due to the series of T-dualities
there is a non-zero component of the B-field
\begin{equation}
B_{\phi_1 \phi_2} = \tilde{\gamma} G \sin^2\theta\cos^2\theta
\end{equation}
We will work in conformal gauge and thus use the Polyakov action
($T=\frac{\sqrt{\lambda}}{2\pi}$)
\begin{multline}
S = \ds\frac{T}{2}\int d^2\sigma [ -(\partial_{\tau}t)^2 +
(\partial_{\tau}\theta)^2 - (\partial_{\sigma}\theta)^2 +
G\sin^2\theta ((\partial_{\tau}\phi_1)^2 -
(\partial_{\sigma}\phi_1)^2) +G\cos^2\theta
((\partial_{\tau}\phi_2)^2 - (\partial_{\sigma}\phi_2)^2)\\ +
2\gamma G\sin^2\theta\cos^2\theta
(\partial_{\tau}\phi_1\partial_{\sigma}\phi_2-\partial_{\sigma}\phi_1\partial_{\tau}\phi_2)]
\end{multline}
which is supplemented by the Virasoro constraints
\begin{equation}
g_{\mu\nu}\partial_{\tau}X^{\mu}\partial_{\sigma}X^{\nu} = 0
\qquad\qquad\qquad
g_{\mu\nu}(\partial_{\tau}X^{\mu}\partial_{\tau}X^{\nu}+\partial_{\sigma}X^{\mu}\partial_{\sigma}X^{\nu})
= 0.
\end{equation}
Here $g_{\mu\nu}$ is the metric (\ref{deformedmetric}) and
$X^{\mu}=\{t,\theta,\phi_1,\phi_2\}$. The ansatz
\begin{equation}
t = \kappa\tau \qquad\qquad \theta = \theta(y) \qquad\qquad \phi_1
= \omega_1\tau + \tilde{\phi}_1(y) \qquad\qquad \phi_2  =
\omega_2\tau + \tilde{\phi}_2(y)
\end{equation}
describes the motion of rigid strings on the deformed 3-sphere,
here we have defined a new variable $y=\alpha\sigma+\beta\tau$.
One can substitute the above ansatz in the equations of motion and
use one of the Virasoro constraints to find three first order
differential equations for the unknown functions:
\begin{equation}
\begin{array}{l}
\tilde{\phi_1}' = \ds\frac{1}{\alpha^2 - \beta^2}\left(
\ds\frac{A}{G \sin^2\theta} +\beta\omega_1
-\tilde{\gamma}\alpha\omega_2\cos^2\theta  \right)\\\\
\tilde{\phi_2}' = \ds\frac{1}{\alpha^2 - \beta^2}\left(
\ds\frac{B}{G\sin^2\theta} +\beta\omega_2
+\tilde{\gamma}\alpha\omega_1\sin^2\theta \right)\\\\
(\theta')^2 = \ds\frac{1}{(\alpha^2-\beta^2)^2} [
(\alpha^2+\beta^2)\kappa^2 - \ds\frac{A^2}{G\sin^2\theta} -
\ds\frac{B^2}{G\cos^2\theta}
-\alpha^2\omega_1^2\sin^2\theta-\alpha^2\omega_2^2\cos^2\theta \\\\
+ 2\tilde{\gamma}\alpha
(\omega_2A\cos^2\theta-\omega_1B\sin^2\theta ) ]
\end{array}
\end{equation}
$A$ and $B$ are integration constants and the prime denotes
derivative with respect to $y$. The other Virasoro constraints
provides the following relation between the parameters
\begin{equation}
A \omega_1 + B \omega_2 + \beta\kappa^2=0
\end{equation}
This system has three conserved quantities - the energy and two
angular momenta:
\begin{equation}
\begin{array}{l}
E = 2 T \ds\frac{\kappa}{\alpha} \int_{\theta_0}^{\theta_1}
\ds\frac{d\theta}{\theta'}\\\\
J_1 = 2  \ds\frac{T}{\alpha} \int_{\theta_0}^{\theta_1}
\ds\frac{d\theta}{\theta'} G\sin^2\theta [\omega_1
+\beta\tilde{\phi}_1' +
\tilde{\gamma}\alpha\cos^2\theta\tilde{\phi}_2']\\\\
J_2 = 2  \ds\frac{T}{\alpha} \int_{\theta_0}^{\theta_1}
\ds\frac{d\theta}{\theta'} G\cos^2\theta [\omega_2
+\beta\tilde{\phi}_2' +
\tilde{\gamma}\alpha\sin^2\theta\tilde{\phi}_1']
\end{array}\label{conserved}
\end{equation}
where the integration is performed over the range of the
coordinate $\theta$. In the analysis below we will find solutions
of the above equations and relations between the energy and the
angular momenta for some special values of the parameters. These
solutions include the giant magnon and the single spike solution
on the deformed $S^3$.

\section{Spikes and giant magnons on $S^2$ and $S^3$}

\subsection{$S^2$ case}

\begin{figure}
\centering
\centerline{ {\includegraphics[height=5cm]{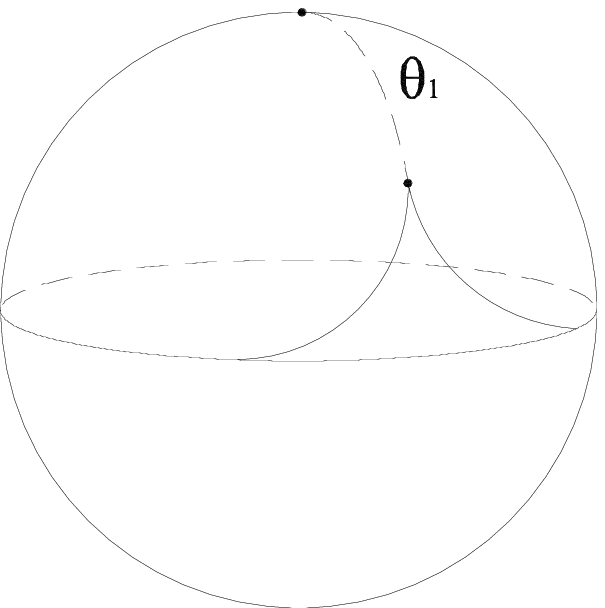} \hskip 2cm
\includegraphics[height=5cm]{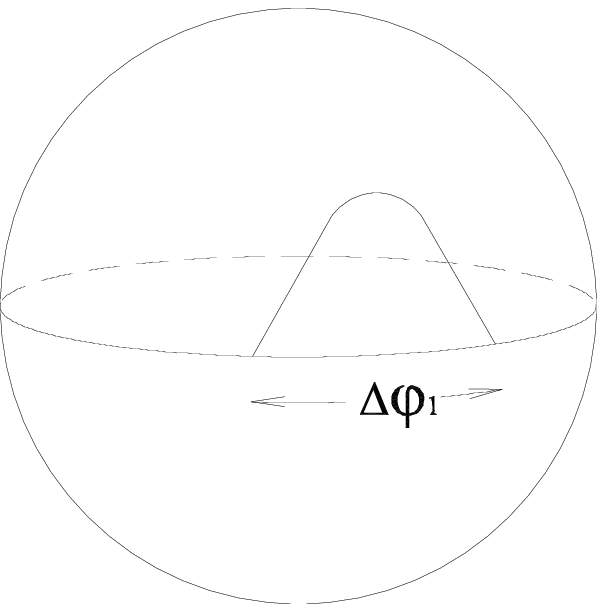}}}
\caption{ A single spike (left) and a giant magnon (right). The
single spike is infinitely wound around the equator and has height
$\frac{\pi}{2}-\theta_1$, whereas the giant magnon has a finite
deficit angle $\Delta\phi_1$, which is interpreted as the momentum
of the corresponding spin chain excitation.}
\label{spike-magnon}
\end{figure}

We can reproduce the known solutions describing a giant magnon
\cite{Hofman:2006xt} and a single spike \cite{Ishizeki:2007we} on
the non-deformed $S^2$ by taking $\tilde{\gamma}=0$ and $\phi_2 =
0$. This simplifies the equations of motion to
\begin{equation}
\begin{array}{l}
\tilde{\phi}_1' =
\ds\frac{1}{\alpha^2-\beta^2}\left(\ds\frac{A}{\sin^2\theta}+\beta\omega_1\right)\\\\
\theta' =
\ds\frac{\alpha\omega_1}{(\alpha^2-\beta^2)\sin\theta}\sqrt{(\sin^2\theta-\sin^2\theta_1)(\sin^2\theta_0-\sin^2\theta)}
\end{array}
\end{equation}
where
\begin{equation}
\sin\theta_0 = \ds\frac{\beta\kappa}{\alpha\omega_1}
\qquad\qquad\qquad \sin\theta_1 = \ds\frac{\kappa}{\omega_1}
\end{equation}
and $\theta$ takes values between $\theta_0$ and $\theta_1$. We
will be interested in solutions describing a string wound around
the equator (Figure \ref{spike-magnon}) and thus require that one
of the turning points $\theta_0$  or $\theta_1$ is equal to
$\pi/2$. This leads to two possible string solutions which, as we
will show, correspond to the giant magnon and the single spike
solutions:
\begin{equation}
\begin{array}{c}
(i)\qquad \ds\frac{\kappa^2}{\omega_1^2} = 1 \qquad\qquad
\text{the giant
magnon solution of \cite{Hofman:2006xt}}\\\\
(ii) \qquad\ds\frac{\kappa^2 \beta^2}{\alpha^2 \omega_1^2} = 1
\qquad\qquad \text{the single spike solution of
\cite{Ishizeki:2007we}}
\end{array}
\end{equation}
Let us first consider case $(i)$, this will give the giant magnon
solution on $S^2$. Using the equation of motion for
$\tilde{\phi}_1$ and the expressions for the energy and the
angular momentum (\ref{conserved}), one finds
\begin{equation}
\begin{array}{l}
E - J_1 = 2 T\cos\theta_0\\\\
\Delta\phi_1 = 2 \ds\int_{\theta_0}^{\frac{\pi}{2}}
\ds\frac{d\theta}{\theta'} \tilde{\phi}_1' =
2\arcsin(\cos\theta_0) = \pi - 2\theta_0
\end{array}
\end{equation}
and thus we end up with the dispersion relation for the giant
magnon on $S^2$
\begin{equation}
E - J_1 = 2 T \cos\theta_0 =
\ds\frac{\sqrt{\lambda}}{\pi}\sin\left(\frac{\Delta\phi_1}{2}\right)
\end{equation}
The angle difference $\Delta\phi_1 $ can be interpreted as the
string counterpart of the momentum $p$ of the magnon excitations
of the corresponding spin chain. The above dispersion relation
matches the one from the spin chain analysis
\cite{Beisert:2005tm}.

Now consider case $(ii)$, this corresponds to the single spike
solution on $S^2$ \cite{Ishizeki:2007we}. The equations of motion
yield the following relations
\begin{equation}
\begin{array}{l}
J_1 = 2T\cos\theta_1 =  \ds\frac{\sqrt{\lambda}}{\pi} \cos\theta_1\\\\
E - T\Delta\phi_1 =
\ds\frac{\sqrt{\lambda}}{\pi}\left(\ds\frac{\pi}{2}-\theta_1\right)
\end{array}
\end{equation}
This solution corresponds to an infinitely wound string around the
equator of $S^2$ (Figure \ref{spike-magnon}) with infinite energy
and finite angular momentum. The interpretation from the field
theory side is not clear but there is a spin chain interpretation
of this system \cite{Ishizeki:2007we}. The single spike can be
thought of as a perturbation of the "slow stings" of
\cite{Roiban:2006jt}

\subsection{$S^3$ case}

If we take $\tilde{\gamma}=0$, we end up with the dynamics of
rigid strings on non-deformed $S^3$. The equations of motion
reduce to
\begin{equation}
\begin{array}{l}
\tilde{\phi}_1' =
\ds\frac{1}{\alpha^2-\beta^2}\left(\ds\frac{A}{\sin^2\theta}+\beta\omega_1\right)\\\\
\tilde{\phi}_2'=
\ds\frac{1}{\alpha^2-\beta^2}\left(\ds\frac{B}{\cos^2\theta}+\beta\omega_2\right)\\\\
\theta' =
\ds\frac{1}{\alpha^2-\beta^2}\sqrt{(\alpha^2+\beta^2)\kappa^2 -
\ds\frac{A^2}{\sin^2\theta} - \ds\frac{B^2}{\cos^2\theta} -
\alpha^2\omega_1^2\sin^2\theta - \alpha^2\omega_2^2\cos^2\theta}
\end{array}\label{eomS3}
\end{equation}
We are interested in string configurations wound around the
equator so we should impose one of the turning points of $\theta$
to be at $\pi/2$. This condition requires $B=0$ and one of the
following two choices
\begin{equation}
\begin{array}{c}
(i)\qquad \ds\frac{\kappa^2}{\omega_1^2} = 1 \qquad\qquad
\text{giant
magnon solution on $S^3$}\\\\
(ii) \qquad\ds\frac{\kappa^2 \beta^2}{\alpha^2 \omega_1^2} = 1
\qquad\qquad \text{single spike solution on $S^3$}
\end{array}
\end{equation}
Let us firs examine the giant magnon solution, i.e.
$\ds\frac{\kappa^2}{\omega_1^2} = 1$ . One of the turning points
of $\theta$ is $\theta_1 = \pi/2$ the other one occurs at
$\sin\theta_0 =
\ds\frac{\beta\omega_1}{\alpha\sqrt{\omega_1^2-\omega_2^2}}$. The
equation of motion for $\theta$ reads
\begin{equation}
\theta' =
\ds\frac{\alpha\sqrt{\omega_1^2-\omega_2^2}}{\alpha^2-\beta^2}\ds\frac{\cos\theta}{\sin\theta}\sqrt{\sin^2\theta
- \sin^2\theta_0}
\end{equation}
This form of $\theta'$ allows one to compute explicitly the
integrals for the conserved quantities (\ref{conserved}) and
obtain the following relations
\begin{equation}
\begin{array}{l}
\Delta\phi_1 = \pi - 2 \theta_0\\\\
E - J_1 = \ds\frac{2 T
\omega_1}{\sqrt{\omega_1^2-\omega_1^2}}\cos\theta_0\\\\
J_2 = \ds\frac{2 T
\omega_2}{\sqrt{\omega_1^2-\omega_2^2}}\cos\theta_0\\\\
\end{array}
\end{equation}
Where to compute these quantities we have used the following
integrals
\begin{equation}
\begin{array}{l}
\ds\int_{\xi}^{\frac{\pi}{2}}
\ds\frac{\sin\theta\cos\theta}{\sqrt{\sin^2\theta - \sin^2\xi}}
d\theta = \cos\xi\\\\
\ds\int_{\xi}^{\frac{\pi}{2}}
\ds\frac{\cos\theta}{\sin\theta\sqrt{\sin^2\theta - \sin^2\xi}}
d\theta = \ds\frac{1}{\sin\xi}\left(\xi - \ds\frac{\pi}{2}\right)
\end{array}
\end{equation}
With the above expressions for the energy and the angular momenta
in hand we find the dispersion realtion of the giant magnon with
two angular momenta on $S^3$.
\begin{equation}
E-J_1 = \sqrt{J_2^2 + \ds\frac{\lambda}{\pi^2}\cos^2\theta_0}
\end{equation}
If we identify the momentum of the spin chain magnon excitation as
$p = \pi - 2\theta_0$ we reproduce the known dispersion relation
for bound states of spin chain magnons
\begin{equation}
E-J_1 = \sqrt{J_2^2 +
\ds\frac{\lambda}{\pi^2}\sin^2\left(\frac{p}{2}\right)}
\end{equation}

The other possible choice of the parameters which will describe a
string wound around the equator is $(ii)$. For this choice we have
$\kappa^2\beta^2=\alpha^2\omega_1^2$ and the equation of motion
for $\theta$ is
\begin{equation}
\theta' =
\ds\frac{\alpha\sqrt{\omega_1^2-\omega_2^2}}{\alpha^2-\beta^2}\ds\frac{\cos\theta}{\sin\theta}\sqrt{\sin^2\theta
- \sin^2\theta_1}.
\end{equation}
where we have defined $\sin\theta_1 =
\ds\frac{\alpha\omega_1}{\beta\sqrt{\omega_1^2-\omega_2^2}}$.
Using this expression it is not hard to find the following
relations between the conserved quantities
\begin{equation}
\begin{array}{l}
J_1 = \sqrt{J_2^2 + \ds\frac{\lambda}{\pi^2} \cos^2\theta_1}\\\\
E - T\Delta\phi_1 =
\ds\frac{\sqrt{\lambda}}{\pi}\left(\ds\frac{\pi}{2}-\theta_1\right)
\end{array}
\end{equation}
which matches the results for the single spike solution first
found in \cite{Ishizeki:2007we} with the Nambu-Goto action. It is
interesting to see how the giant magnons and single spike
solutions arise as limiting cases of a rigid string rotating
around the equator, this implies that they fall into the same
class of solutions and suggests that the single spike solutions
might be important from the spin chain/gauge theory point of view.

\section{Relation to the sine-Gordon  model}

The classical string solutions on $S^3$ described in the previous
section have a relation to the sine-Gordon model. This is
important for calculations of the scattering of single spikes and
magnons. This relation can be explicitly seen by computing the
determinant of the two dimensional world-sheet metric
\begin{equation}
\sqrt{-h} = \alpha^2 (\theta'^2 + \sin^2\theta\tilde{\phi}_1'^2 +
\cos^2\theta\tilde{\phi}_2'^2)
\end{equation}
where we have used the Virasoro constraints. For the giant magnon
solution we have $\kappa^2=\omega_1^2$ and we can use the integral
\begin{equation}
\ds\int \frac{\sin\theta d\theta}{\cos\theta\sqrt{\sin^2\theta -
\sin^2\xi}} = \ds\frac{1}{\cos\xi}
\text{arctanh}\left(\sqrt{\ds\frac{\cos(2\xi) - \cos(2\theta)}{1 +
\cos(2\xi)}}\right)
\end{equation}
to find the following solution for $\theta$:
\begin{equation}
\cos\theta = \ds\frac{\cos\theta_0}{\cosh(C y)} \qquad
\text{where} \qquad C =
\ds\frac{\alpha\sqrt{\omega_1^2-\omega_2^2}}{\alpha^2-\beta^2}\cos\theta_0
\end{equation}
and thus
\begin{equation}
\ds\frac{\sqrt{-h}}{(\alpha^2-\beta^2)C^2} =
\ds\frac{1}{\cosh^2(Cy)}
\end{equation}
If we define the sine-Gordon field as $\sin^2\Psi_0 =
\ds\frac{\sqrt{-h}}{(\alpha^2-\beta^2)C^2} =
\ds\frac{1}{\cosh^2(Cy)}$ it is straightforward to check that
$\Psi_0$ is a solution to the sine-Gordon equation
\begin{equation}
\partial_{\tau}^2\Psi_0 - \partial_{\sigma}^2\Psi_0 +
\ds\frac{(\alpha^2-\beta^2)C^2}{2}\sin(2\Psi_0)=0
\end{equation}
One can perform analogous calculation for the single spike
solution on $S^3$ with $\beta^2\kappa^2=\alpha^2\omega_1^2$. For
this case we have a solution for $\theta$ of the form
\begin{equation}
\cos\theta = \ds\frac{\cos\theta_1}{\cosh(D y)} \qquad
\text{where} \qquad D =
\ds\frac{\alpha\sqrt{\omega_1^2-\omega_2^2}}{\alpha^2-\beta^2}\cos\theta_1
\end{equation}
and if we impose $D^2 = \ds\frac{\kappa^2}{\beta^2-\alpha^2}$ we
find
\begin{equation}
\sin^2\Psi_1 = \ds\frac{\sqrt{-h}}{(\beta^2-\alpha^2)D^2} =
\tanh^2(Dy)
\end{equation}
Where $\Psi_1$ solves the sine-Gordon equation
\begin{equation}
\partial_{\tau}^2\Psi_1 - \partial_{\sigma}^2\Psi_1 +
\ds\frac{(\beta^2-\alpha^2)D^2}{2}\sin(2\Psi_1)=0
\end{equation}
Although we do not explore further the connection between the
single spike solutions on $S^3$ and the sine-Gordon model it seems
interesting to use it and study scattering of single spikes as
this was done for the giant magnons.

\section{$\gamma$-deformations, spiky strings and giant magnons}

In this section we will look for string solutions wound around the
equator of the $\gamma$-deformed $S^3$. To ensure this we again
require that one of the turning points of $\theta$ occurs at
$\theta=\pi/2$. This imposes some constraints on the parameters of
the solution. It turns out that the deformation parameter does not
play a role on the constraints and they are the same as for the
undeformed case, namely $B=0$ and
\begin{equation}
\kappa^2 = \omega_1^2 \qquad\qquad \text{or} \qquad\qquad
\beta^2\kappa^2 = \alpha^2\omega_1^2
\end{equation}

As was noted in \cite{Chu:2006ae,
Bobev:2006fg}\footnote{Semiclassical strings solutions in the
Lunin-Maldacena background were considered in
\cite{froits},\cite{tseytfrroi},\cite{bodira1},\cite{Ryang:2005pg},\cite{Chen:2006bh}
see also \cite{McLoughlin:2006cg}, for a nice review on the
subject and a complete list of references see
\cite{Swanson:2007dh}}, the non-trivial deformation of the $\axs$
background imposes certain conditions. It turns out that this
solution is required to live on the ``three-sphere'' and we need
to consider a classical string moving on $S^3_{\gamma}$. As before
we, use (\ref{3-sphere}) to parametrize the ambient space, i.e.
the deformed three-sphere  is parameterized by the three angles
$\theta, \phi_1$ and $\phi_2$.
\subsection{Giant magnons}

If we choose $\kappa^2 = \omega_1^2$ (which through the Virasoro
constraint implies $A = -\beta\omega_1$) we get the giant magnon
solution on $S^3_{\gamma}$ found in \cite{Bobev:2006fg}. The
equations of motion for this case are:
\begin{equation}
\begin{array}{l}
\tilde{\phi}_1' = - \ds\frac{\cos^2\theta}{\alpha^2-\beta^2}
\left(\ds\frac{\beta\omega_1}{\sin^2\theta} +
\tilde{\gamma}\alpha\omega_2+\tilde{\gamma}^2\beta\omega_1 \right)\\\\
\tilde{\phi}_2' = \ds\frac{\beta\omega_2 +
\tilde{\gamma}\alpha\omega_1\sin^2\theta}{\alpha^2 - \beta^2}\\\\
\theta' =
\ds\frac{\alpha\Omega_0}{(\alpha^2-\beta^2)}\ds\frac{\cos\theta}{\sin\theta}\sqrt{\sin^2\theta
-\sin^2\theta_0}
\end{array}\label{eomdefmagnon}
\end{equation}
where we have defined
\begin{equation}
\sin\theta_0 = \ds\frac{\beta\omega_1}{\alpha\Omega_0}
\qquad\qquad \text{and} \qquad\qquad \Omega_0 =
\ds\sqrt{\omega_1^2 - \left(\omega_2 +
\tilde{\gamma}\ds\frac{\beta\omega_1}{\alpha}\right)^2}
\end{equation}
Using the expressions for the energy and the angular momentum
(\ref{conserved}) and equations (\ref{eomdefmagnon}) we find
\begin{equation}
\begin{array}{l}
E - J_1 = 2T\ds\frac{\omega_1}{\Omega_0}\cos\theta_0\\\\
J_2 = 2T\left(\ds\frac{\omega_2}{\Omega_0} +
\tilde{\gamma}\ds\frac{\beta\omega_1}{\alpha\Omega_0}\right)\cos\theta_0
\end{array}
\end{equation}
These expressions lead to the dispersion relation for the giant
magnon solution on $\gamma$-deformed $S^3$ \cite{Chu:2006ae},
\cite{Bobev:2006fg}
\begin{equation}
E-J_1 = \ds\sqrt{J_2^2 + \ds\frac{\lambda}{\pi^2}\cos^2\theta_0}
\end{equation}
In order to make a connection with the spin chain description we
should identify $\cos\theta_0 =
\sin\left(\frac{p}{2}-\pi\beta\right)$, where $p$ is the momentum
of the magnon excitation on the spin chain and $\beta =
\tilde{\gamma}/\sqrt{\lambda}$. So the prediction for the relevant
spin chain dispersion relation is
\begin{equation}
E-J_1 = \ds\sqrt{J_2^2 +
\ds\frac{\lambda}{\pi^2}\sin^2\left(\frac{p}{2}-\pi\beta\right)}
\end{equation}
this relation is invariant under $p\rightarrow p + 2\pi$ and
$\beta\rightarrow \beta+1$ as is required by the spin chain
analysis \cite{froits}, \cite{Beisert:2005if}. We note that in the
deformed background we find
\begin{equation}
\ds\frac{\Delta\phi_1}{2} = \ds\left(\frac{\pi}{2} -
\theta_0\right) - \tilde{\gamma}\ds\frac{\sqrt{\omega_1^2 -
\Omega_0^2}}{\Omega_0}\cos\theta_0
\end{equation}
and in contrast to the undeformed giant magnon solution the
momentum $p$ of the spin chain magnon is no longer interpreted as
the angular difference $\Delta\phi$.

\subsection{Single spikes}

Considering the case $\beta^2\kappa^2 = \alpha^2\omega_1^2$, which
yields $A = -\frac{\omega_1\alpha^2}{\beta}$, we expect to find a
rigid string solution solution on $S^3_{\gamma}$ which is the
analogue of the single spike solution on $S^3$ found in
\cite{Ishizeki:2007we}. The equations of motion are
\begin{equation}
\begin{array}{l}
\tilde{\phi}_1' = \ds\frac{1}{\alpha^2-\beta^2}\left(
\beta\omega_1 - \ds\frac{\alpha^2\omega_1}{\beta\sin^2\theta} -
\tilde{\gamma}\alpha\sqrt{\omega_1^2-\Omega_1^2}\cos^2\theta
\right)\\\\
\tilde{\phi}_1' = \ds\frac{1}{\alpha^2-\beta^2}(\beta\omega_2 +
\tilde{\gamma}\alpha\omega_1\sin^2\theta)\\\\
\theta' =
\ds\frac{\alpha\Omega_1}{(\alpha^2-\beta^2)}\ds\frac{\cos\theta}{\sin\theta}\sqrt{\sin^2\theta
-\sin^2\theta_1}
\end{array}
\end{equation}
where
\begin{equation}
\sin\theta_1 = \ds\frac{\alpha\omega_1}{\beta\Omega_1}
\qquad\qquad\qquad \Omega_1 = \ds\sqrt{\omega_1^2 - \left(\omega_2
+ \tilde{\gamma}\ds\frac{\alpha\omega_1}{\beta}\right)^2}
\end{equation}
The two angular momenta are
\begin{equation}
J_1 = 2 T\ds\frac{\omega_1}{\Omega_1}\cos\theta_1 \qquad\qquad J_2
= -2T\ds\frac{\sqrt{\omega_1^2-\Omega_1^2}}{\Omega_1}\cos\theta_1
\end{equation}
This leads to the relation
\begin{equation}
J_1 = \sqrt{J_2^2 + \ds\frac{\lambda}{\pi^2} \cos^2\theta_1}
\end{equation}
This looks identical to the corresponding expression in the
undeformed case, the dependence on the deformation parameter
$\tilde{\gamma}$ is buried in the definition of $\cos\theta_1$.
Similarly to the giant magnon solution if we want to make a
connection with the spin chain description we should identify
$\cos\theta_1 = \sin\left(\frac{p}{2}-\pi\beta\right)$. However
the interpretation of the single spike solutions from the dual
spin chain/gauge theory side is not completely clear.

For the relation between $E$ and $\Delta\phi_1$ we find:
\begin{equation}
E - T\Delta\phi_1 =
\ds\frac{\sqrt{\lambda}}{\pi}\left(\ds\frac{\pi}{2}-\theta_1\right)
-
\tilde{\gamma}\ds\frac{\sqrt{\lambda}}{\pi}\ds\frac{\sqrt{\omega_1^2-\Omega_1^2}}{\Omega_1}\cos{\theta_1}
\end{equation}
As should be expected in the limit $\tilde{\gamma}\rightarrow 0$
this expression reduces to the one for the single spike solution
on undeformed $S^3$. Similarly to the deformed giant magnon
solution we see that there is a non-trivial $\tilde{\gamma}$
dependence in the angle difference $\Delta\phi_1$ which prevents
us from interpreting $E - T\Delta\phi_1$ as the momentum of the
spin chain excitation.

\section{Comments and conclusions}

In this short paper we investigated the multispin string solutions
representing infinitely wound strings which develop a single spike
in the Lunin-Maldacena background \cite{Lunin:2005jy}. First we
gave a derivation of the solutions in conformal gauge using the
Polyakov sigma model action on $S^3_{\gamma}$. When reduced to the
undeformed $S^2$ and $S^3$ the solutions actually reproduce the
single spike, or giant magnon solutions of \cite{Hofman:2006xt},
\cite{Ishizeki:2007we}. Although the results are the same, the
derivation is performed with the sigma model action which allows
to make connections to more general cases, for instance those
based on the reduction to the Neumann-Rosochatius system
\cite{Kruczenski:2006pk} or to solutions on deformed backgrounds.
The new solution we found is a single spike solution on the
deformed $S^3$ of the Lunin-Maldacena background, it has two
finite angular momenta and infinite energy and winding angle.

In contrast to the magnon solutions, in the single spike case the
role of the angle  difference $\Delta \phi_1$  and the spin $J_1$
interchange. The angle difference becomes infinite while all the
spins $J_i$ remain finite. Nevertheles, the difference $E-\Delta
\phi_1$ is finite. Formally one can interpret the angle $\pi-2
\theta_1$ as the momentum of the corresponding spin chain
excitation making the expressions very similar to the ones for the
giant magnon. However in this case one should interpret $J_1$ and
not $E$ as the energy of the excitation
$$
E-T\Delta
\phi=\frac{\sqrt{\l}}{\pi}\left(\ds\frac{\pi}{2}-\theta_1\right)
$$
Then one can loosely think of this case as arising from an
interchange of the worldsheet coordinates $\sigma$ and $\tau$.
This is supported also by the way the single spike solution
reduces to the sine-Gordon model. In the case of dual spikes in
AdS part of geometry the interpretation is somehow more
transparent but we didn't considered that case here\footnote{For
discussion on this point see \cite{Mosaffa:2007}.}.

The new single spike solution we obtained in the gamma deformed
geometry share some of the properties of the deformed giant
magnons. The angle difference $\Delta\phi_1$ is shifted by a
$\tilde\gamma$-dependent term. For the deformed magnon solution it
is given by \eq{ \frac{\Delta\phi}{2}=\ds\left(\frac{\pi}{2} -
\theta_0\right) - \tilde{\gamma}\ds\frac{\sqrt{\omega_1^2 -
\Omega_0^2}}{\Omega_0}\cos\theta_0 \notag } where
$$
\Omega_0 = \ds\sqrt{\omega_1^2 - \left(\omega_2 +
\tilde{\gamma}\ds\frac{\beta\omega_1}{\alpha}\right)^2}.
$$
and for the deformed single spike we find \eq{ E - T\Delta\phi =
\ds\frac{\sqrt{\lambda}}{\pi}\left(\ds\frac{\pi}{2}-\theta_1\right)
-
\tilde{\gamma}\ds\frac{\sqrt{\lambda}}{\pi}\ds\frac{\sqrt{\omega_1^2-\Omega_1^2}}{\Omega_1}\cos{\theta_1}
\notag } where
$$
\Omega_1 = \ds\sqrt{\omega_1^2 - \left(\omega_2 +
\tilde{\gamma}\ds\frac{\alpha\omega_1}{\beta}\right)^2}
$$
In both cases $\Omega_i$ determine the turning points of the
string $\theta_0$ and $\theta_1$ \al{ &
\sin\theta_0=\frac{\beta\omega_1}{\a\Omega_0} \quad \text{magnon
case} \notag \\
& \sin\theta_1=\frac{\beta\omega_1}{\a\Omega_1} \quad \text{single
spike case}. \notag } The relation between the angular momenta of
the deformed single spike solution looks the same as the one for
the undeformed case
\begin{equation}
J_1 = \sqrt{J_2^2 + \ds\frac{\lambda}{\pi^2} \cos^2\theta_1}
\end{equation}
however there is implicit dependence on the deformation parameter
$\tilde{\gamma}$ in the definition of $\theta_1$.

The study of the magnon and single spike solutions suggests
several interesting directions  for further studies. First of all
the interpretation of the single spike solutions in terms of
mapping to a spin chain is still not quite clear especially on the
field theory side. There could be a relation between the single
spike solutions and the ``slow strings" of \cite{Roiban:2006jt}
and a possible interpretation of the single spikes as excitation
of the "antiferromagnetic" state of the spin chain. Studying the
scattering of single spikes and comparison with the magnon case
can shed light on the interpretation of these solutions. The
single spike solutions could be generalized to spikes on $S^5$ as
well as to string solutions with dynamics on the full $\axs$
geometry. It would be interesting to find also spiky solutions
which are not rigid but are both rotating and pulsating. We hope
to report on some of these issues in the near future.

\bigskip
\leftline{\bf Acknowledgements}
\smallskip

It is a pleasure to thank Martin Kruczenski for reading the draft
of the paper and helpful correspondence. R.R. acknowledges a Guest
Professorship and warm hospitality at the Institute for
Theoretical Physics, Vienna University of Technology. Special
thanks to Max Kreuzer and his group for fruitful discussions,
asking many questions and stimulating atmosphere. N.P.B. is
grateful to Nick Warner for his constant encouragement and
support. The work of N.P.B. was supported in part by funds
provided by the DOE under grant DE-FG03-84ER-40168 and by a Dean
Joan M. Schaefer Research Scholarship. The work of R.R is
partially supported by Austrian Research Fund FWF grant \#
P19051-N16 and Bulgarian NSF BUF-14/06 grant.

\section*{Appendix A. Single spikes from the Neumann-Rosochatius
integrable system}
\appendix
\renewcommand{\theequation}{A.\arabic{equation}}
\setcounter{equation}{0}  

In this appendix we will derive single spike solutions in $S^3$
using reduction to the Neumann-Rosochatius (NR) integrable system
\cite{Kruczenski:2006pk}. The reduction to the NR system can be
done using the following generalized ansatz for the five-sphere
coordinates
\eq{ X_a=x_a(y)e^{i\omega_a\tau},\,\, a=1,2,3; \quad
y=\a\sigma+\b\tau,}
where the complex quantities $x_a$ satisfy the periodicity
condition
\eq{x_a(y+2\pi\a)=x_a(y), \quad x_a=r_a(y)e^{i\m_a(y)}\,\,
\,\,\text{($r_a$ real)} \label{param-sph} }
Substituting in the Polyakov action one can find
\ml{\L=\sum\limits_a
\lb[(\a^2-\b^2){r'}_a^2+(\a^2-\b^2)r^2_a\lb({\m'}_a-\frac{\b\omega_a}{\a^2-\b^2}\rb)^2
-\frac{\a^2}{\a^2-\b^2}\omega_a^2r_a^2\rb] \\
+\Lambda\lb(\sum\limits_a r^2_a-1\rb). }
One can find $\m_a$ from the equation
\eq{\m'_a=\frac{1}{\a^2-\b^2}\lb[\frac{C_a}{r_a^2}+\b\omega_a\rb]}
where $C_a$ are integration constants. The the Lagrangian reduces
to
\eq{\L=\sum\limits_a\lb[(\a^2-\b^2){r'}_a^2-\frac{C_a}{(\a^2-\b^2)r_a^2}-\frac{\a^2}{\a^2-\b^2}\omega_a^2r_a^2\rb]
+\Lambda\lb(\sum\limits_a r^2_a-1\rb).}
The conserved quantities are given by the  expressions
\al{& E=T\frac{\kappa}{\a}\int dy \label{energ} \\
& J_a=T\int
dy\lb(\frac{\b}{\a}\frac{C_a}{\a^2-\b^2}+\frac{\a}{\a^2-\b^2}\omega_ar_a^2 \rb) \label{spins} \\
& \left(\frac{\a^2+\b\sum\limits_a
C_a/\omega_a}{\a^2-\b^2}\right)\frac{E}{\kappa}=
\sum\limits_a\frac{J_a}{\omega_a} }
where the last relation comes from the constraints.

The reduction to the $S^3$ case implies $r_3=\m_3=0$ (also
$C_3=0$). In addition, the condition that one of the turning
points of the string is at $\theta=\pi/2$ imposes $ C_2=0$,
 $\b=-\frac{\omega_1C_1}{\kappa^2}$ (we set $\a=1$ and then $|\b|$
becomes a ''group velocity``). The equation for $\kappa$ has two
solutions: $\kappa=\omega_1$ corresponding to a magnon type
solution, and $\kappa= C_1$ corresponding to single spike
solution.

The conserved quantities are (see eqs (3.7)-(3.10) in
\cite{Kruczenski:2006pk})
\al{& \hat\m=\frac{C_1}{(1-\b^2)}\int\frac{dy}{u}
+\frac{\beta\omega_1}{(1-\beta^2)}\int dy \notag \\
& E=\kappa T\int dy \notag \\
& J_1=\frac{C_1\beta}{(1-\b^2)}\int dy +
\frac{\omega_1}{(1-\b^2)}T\int
u\,dy \\
& J_2=\frac{\omega_2}{(1-\b^2)}T\int(1-u)dy \notag }
where $\hat\m_1$ corresponds to our $\phi_1$ and
$u=r_1^2=\sin^2\theta$.

To find finite results (which is so for $E-J_1$ in magnon case) we
consider
\eq{E-T\hat\m_1=\frac{2C_1T}{\sqrt{\Delta\omega^2}}\arccos\sqrt{\bar
u}=\frac{\sqrt{\lambda}}{\pi}\bar\theta}
where
\eq{\bar\theta=\frac{\pi}{2}-\theta_0}
For the angular momenta we get
\eq{J_1=\frac{2T\omega_1}{\sqrt{\Delta\omega^2}}\cos\theta_0=\frac{2T\omega_1}{\sqrt{\Delta\omega^2}}\sin\bar\theta}
Analogously
\eq{J_2=-\frac{2T\omega_2}{\sqrt{\Delta\omega^2}}\cos\theta_0=-\frac{2T\omega_2}{\sqrt{\Delta\omega^2}}\sin\bar\theta}
Defining $\sin\gamma$ as in \cite{Ishizeki:2007we}
\eq{ \sin\gamma=\frac{\omega_2}{\omega_1}, \qquad\qquad
\sin\theta_0=\frac{C_1}{\sqrt{\Delta\omega^2}}}
we find
\al{& J_1=2T\frac{1}{\sin\gamma}\sin\bar\theta \notag \\
& J_2=-2T\frac{\sin\gamma}{\cos\gamma}\sin\bar\theta}
These give the result for dispersion relations eqs. (6.24)-(6.25)
of \cite{Ishizeki:2007we}
\al{& E-T\hat\m_1=\frac{\sqrt{\lambda}}{\pi}\bar\theta \\
& J_1^2=J_2^2+\frac{\lambda}{\pi^2}\sin^2\bar\theta}
As a result we rederived the results for single spikes in $S^3$
with two spins. We see that setting $J_2=0$ to the single spike
with one angular momentum.

Using ellipsoidal coordinates as in \cite{Kruczenski:2006pk} one
can find for motion in $S^5$ the results
\al{E-T\Delta\hat\m_1=\frac{\sqrt{\lambda}}{\pi}(\bar\theta_2+\bar\theta_3)
\notag \\
J_1=\sqrt{J_2^2+\frac{\l}{\pi^2}\sin\bar\theta_2}+\sqrt{J_3^2+\frac{\l}{\pi^2}\sin\bar\theta_3}
}
The details of this derivation will be given elsewhere
\cite{bora}.

\end{document}